\documentclass[twocolumn,english,aps,prb,superscriptaddress]{revtex4}
\usepackage[T1]{fontenc}
\usepackage[latin1]{inputenc}
\usepackage{amsmath}
\usepackage{amssymb}
\usepackage{color}
\usepackage{graphicx}
\usepackage{amssymb}

\makeatletter

\providecommand{\LyX}{L\kern-.1667em\lower.25em\hbox{Y}\kern-.125emX\@}

\usepackage{graphicx}
\usepackage{color}
\usepackage{babel}
\makeatother
\begin{document}

\author{A. Friedrich}

\email{afried@physik.rwth-aachen.de}

\affiliation{RWTH Aachen, Institut f\"ur Theoretische
Physik C, Physikzentrum, 52056 Aachen}

\author{W. Brenig}

\affiliation{Technische Universit\"at Braunschweig, Institut f\"ur Theoretische
Physik, Mendelssohnstr. 3, 38106 Braunschweig}

\date{\today}

\begin{abstract}
We study the interplay of spin, orbital and lattice degrees of freedom
in a one-dimensional Kugel-Khomskii
model coupled to phonons. In the vicinity of the dimer
point we analyze the excitation spectrum, mapping the
spin and orbital degrees of freedom to bond operators.
In particular we study the renormalization of the phonon
propagator due to coupling to the orbital and magnetic excitations.
Considering both, ferro- and antiferro-orbital ordering we will show
that the appearance of orbiton shake-up processes in the phonon spectrum
is sensitive to the type of Jahn-Teller distortion. The relevance
of our results for optical spectroscopy on low dimensional transition
metal compounds will be discussed. 
\end{abstract}

\title{Lattice Dynamics in a strongly dimerized low-dimensional Model with
Orbital Ordering}

\maketitle

\section{Introduction}

Orbital ordering in three dimensions (3D) is a well established
phenomenon in a variety of transition-metal compounds\cite{elbio02}.
Very recently, quasi one-dimensional (1D) systems with potentially
similar behavior, such as Na$_{2}$Ti$_{2}$Sb$_{2}$O\cite{axtell} and 
NaTiSi$_2$O$_6$\cite{konstantinovic02} have 
attracted considerable interest. In
these systems the coupling of orbital degrees of freedom to those
of the spin and probably also to the lattice have been suggested to
lead to a rich variety of possible phases and non-trivial elementary
excitations\cite{Azaria00}. In principle the orbitally ordered state
will allow for elementary excitations of orbital nature, i.e. \emph{orbitons}.
However, while their existence has been predicted more than two decades
ago\cite{kugel-khomskii} it has been only very recently that an indirect
observation of orbitons in 3D has been claimed in shake-up side-bands
of Raman-active phonon spectra in LaMnO$_{3}$ \cite{saitoh}. This
observation underlines the relevance of a simultaneous description
of orbital and lattice degrees of freedom. Unfortunately however,
for quasi 1D materials this is still an open issue. It is the aim of
this work to investigate the phonon spectra of a 1D model of mixed
orbital and spin degrees of freedom by investigating its harmonic
lattice dynamics in the vicinity of two specifically chosen Jahn-Teller
(JT) distortions. The outline of the paper is as follows. In section
I we discuss the 1D spin-orbital Hamiltonian and its excitations using
a bond-operator method close to the so-called dimer line. Sections
II and III will focus on the phonon-spectra which form in the presence
of a frozen-in, either ferro- or antiferro-orbital JT distortions.
Conclusions will be presented in the final section IV. Technical details
are deferred to appendices A through D.

\section{Bond-Operator Description of the 1D Kugel-Khomskii Model}

In this section, and prior to discussing their effect on the lattice
dynamics we develop a description of the orbital and spin excitations.
In the limit of strong Coulomb repulsion, orbital degeneracy in the
Hubbard model can be described in terms of a pseudo-spin model, as
was shown several decades ago by Kugel and Khomskii\cite{kugel-khomskii}. In 1D
and for a two-fold orbital degeneracy at 1/4-filling this leads to
the Hamiltonian of a magnetic spin-1/2 chain coupled to an orbital
pseudo-spin-1/2 chain by bi-quadratic interactions 
\begin{equation}
\begin{split} H= & J_{1}\sum _{n}S_{n}^{\alpha }S_{n+1}^{\alpha }+J_{2}\sum _{n}\left(T_{n}^{\alpha }T_{n+1}^{\alpha }+AT_{n}^{z}T_{n+1}^{z}\right)\\
 + & K\sum _{n}S_{n}^{\alpha }S_{n+1}^{\alpha }\left(T_{n}^{\gamma }T_{n+1}^{\gamma }+BT_{n}^{z}T_{n+1}^{z}\right)\end{split}
\label{KKchain}\end{equation}
 where, apart from the spin-$1/2$ operator $S_{n}$, a pseudo spin-$1/2$ variable $T_{n}$ with $T_{n}^{z}=\pm 1/2$ corresponding to the orbital degrees of freedom   is attached to each site '$n$'. Summation over the
spin indices $\alpha =x,y,z$ and $\gamma =x,y,z$ is implied. In
the remainder of this paper energies will be given in units of $K$.

The quantum phase diagram of (\ref{KKchain}) has been the subject
of intense research\cite{Pati98,Yamashita98,Frischmuth99,Itoi00,Azaria00,Kolezhuk98_00}.
For $J_{1},J_{2}>1/4$ its spectrum is believed to be massive\cite{Pati98}.
At $J_{1}$=$J_{2}$=1/4 and $A$=$B$=0 the model has SU(4) symmetry
and is integrable\cite{Azaria00}. Along the 'dimer line' $J_{1}$=$(3+A)/4$,
$J_{2}$=3/4 with $-2<A<\infty $ it displays a two-fold degenerate,
fully dimerized ground state. At the 'dimer point' $J_{1}$=$J_{2}$=3/4
this ground state can be expressed as a direct matrix-product of alternating
spin and orbital singlets with an energy gap of $\Delta \sim 0.375$\cite{Pati98,Itoi00,Kolezhuk98_00}.
Exactly on the dimer line the elementary excitations are massive solitons,
i.e. domain walls between the two-fold degenerate ground-states. However,
off from the dimer-line the latter readily bind to form generalized
gapful triplet-modes\cite{Kolezhuk98_00}.

\begin{figure}
\begin{center}\includegraphics[width=0.8\columnwidth,
  keepaspectratio,
  origin=c]{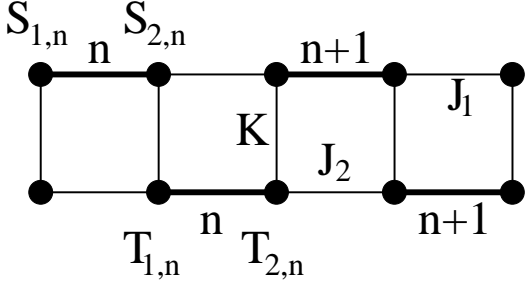}\end{center}
\caption{\label{fig1} Dimer representation of the spin-orbital chain. $n$ labels the dimer, the index $1,2$ the site on a dimer. On each site on the upper (lower) chain sits one spin $S_{i,n}$ (pseudo-spin $T_{i,n}$ i.e.orbital degree of freedom) respectively.}
\end{figure}

Prior to including the effects of a JT-distortion we now reformulate
(\ref{KKchain}) in terms of bosonic bond-operators\cite{sachdev,Starykh,WB}.
This allows for an approximate description of the formation of the
spin and orbital singlets in the ground state of the dimer phase \begin{equation}
\begin{split} S_{_{2}^{1},n}^{\alpha }= & \frac{1}{2}(\pm s_{n}^{\dagger }t_{\alpha ,n}\pm t_{\alpha ,n}^{\dagger }s_{n}-i\epsilon _{\alpha \beta \gamma }t_{\beta ,n}^{\dagger }t_{\gamma ,n})\\
 T_{_{2}^{1},n}^{\alpha }= & \frac{1}{2}(\pm \sigma _{n}^{\dagger }\tau _{\alpha ,n}\pm \tau _{\alpha ,n}^{\dagger }\sigma _{n}-i\epsilon _{\alpha \beta \gamma }\tau _{\beta ,n}^{\dagger }\tau _{\gamma ,n})\end{split}
\label{SinBB}\end{equation}
 where $s(\sigma )_{n}^{(^{\dagger })}$ and $t(\tau )_{n}^{(^{\dagger })}$
create/destroy spin(orbital) singlets and triplets and the index $1(2)$
labels the left (right) site of the dimer $n$ in Fig. \ref{fig1}. 
Bose operators labeled by roman letters refer to spin
space, while greek ones refer to orbital space. The bond-operators
have to satisfy \begin{equation}
s(\sigma )_{n}^{\dagger }s(\sigma )_{n}+t(\tau )_{\alpha ,n}^{\dagger }t(\tau )_{\alpha ,n}=1\label{constr}\end{equation}
 which implies the constraint of no double-occupancy of the bonds by
singlets or triplets.

Inserting (\ref{SinBB}) into (\ref{KKchain}) an interacting Bose
gas is obtained. In the limit of strong dimerization one may proceed
with the latter by invoking the Holstein-Primakoff (LHP) approximation\cite{Starykh,WB}.
I.e. the singlets are removed using (\ref{constr}) and assuming $s^{(\dagger )}$
and $\sigma ^{(\dagger )}$ to be $C$-numbers. This is consistent
with a condensation of the singlets in the ground state, i.e. the
formation of a dimer state. The procedure leads to square roots of
type $s^{(\dagger )}(\sigma ^{(\dagger )})=[1-t(\tau )_{\alpha ,n}^{\dagger }t(\tau )_{\alpha ,n}]^{1/2}$
which, approximately, may be linearized to retain triplet interactions
up to quartic order. Since we are interested in the effects of mutual
interactions between the orbital and magnetic sector we simplify even
further, by keeping those interactions which mix the orbital and magnetic
triplets, but discarding quartic terms of purely orbital or magnetic
nature. After some algebra we find a quadratic part of the Hamiltonian\begin{alignat}{1}
H_{0} & =-\frac{N}{4}\left[3\left(J_{1}+J_{2}\right)+J_{2}A\right]\nonumber \\
 & +\sum _{n,m,i,j}\Phi _{n,i}^{*}M_{n,m}^{ij}\Phi _{m,j}\label{hnull}
\end{alignat}
 with $\Phi _{n}=(t_{\alpha,n}^{\dagger },t_{\alpha,n},\tau _{x,y,n}^{\dagger },\tau _{x,y,n},\tau _{z,n}^{\dagger },\tau _{z,n})$
and again $\alpha=x,y,z$.
$n$ and $m$ label dimers and $i$ and $j$ the components of $\Phi ^{*}$
and $\Phi $. The matrix elements $M_{n,m}^{ij}$ are given in Appendix
\ref{matrixhnull}. We note, that on the quadratic level no mixing
occurs between the spin and orbital triplets. For the quartic spin-orbital
interactions we get
\begin{alignat}{1}
H_{SO} & =-\frac{1}{4}\sum _{n}\left\{ \left[\tau _{z,n}^{\dagger }\tau _{z,n}+\left(1+\frac{B}{2}\right)\tau _{\beta ,n}^{\dagger }\tau _{\beta ,n}\right]\right.\nonumber \\
 & \left.\times \left(t_{\alpha ,n-1}^{\dagger }t_{\alpha ,n}+t_{\alpha ,n-1}^{\dagger }t_{\alpha ,n}^{\dagger }+h.c.\right)\right\} \nonumber \\
 & -\frac{1}{4}(1+B)\sum _{n}\left[t_{\alpha ,n}^{\dagger }t_{\alpha ,n}\times \right.\label{eq:Hso}\\
 & \left.\left(\tau _{z,n-1}^{\dagger }\tau _{z,n}+\tau _{z,n-1}^{\dagger }\tau _{z,n}^{\dagger }+h.c.\right)\right]\nonumber \\
 & -\frac{1}{4}\sum _{n}t_{\alpha ,n}^{\dagger }t_{\alpha ,n}\left(\tau _{\beta ,n-1}^{\dagger }\tau _{\beta ,n}+\tau _{\beta ,n-1}^{\dagger }\tau _{\beta ,n}^{\dagger }+h.c.\right)\, \nonumber 
\end{alignat}
with $\beta = x,y$. The quadratic part $H_{0}$ can be diagonalized by Fourier- and Bogoliubov
transformation leading to dispersions \begin{eqnarray}
H_{LHP} & = & \sum _{k,\alpha ,x}\omega _{x,k}x_{k,\alpha }^{\dagger }x_{k,\alpha }\label{hquad}\\
\omega _{x,k} & = & \chi _{1,x}\left[\chi _{2,x}+\chi _{3,x}\cos \left(k\right)\right]^{\frac{1}{2}}\, \, \, ,\label{lhpdispersion}
\end{eqnarray}
 where $x=a,c$ label new quasi-particles of purely spin ($a$) and
orbital ($c$) nature with $t_{\alpha ,k}^{\dagger }=u_{k}a_{\alpha ,k}^{\dagger }+v_{k}a_{\alpha ,-k}$
and $\tau _{\alpha ,k}^{\dagger }=g_{k}c_{\alpha ,k}^{\dagger }+h_{k}c_{\alpha ,-k}$.
Explicit expressions for the dispersions $\omega _{x,k}$, the Bogoliubov
coefficients, and $\chi _{i,x}$ are listed in Appendix \ref{dispersionen}.
After Fourier-transformation the spin-orbital interaction $H_{SO}$
reads\cite{jurecka}\begin{alignat}{1}
H_{SO} & =\sum _{k,k',q}\left[a_{\alpha ,k'-q}^{\dagger }a_{\alpha ,-k'}^{\dagger }\right.\nonumber \\
 & \left.\times \, \left(V_{k,k',q}^{1,\gamma }c_{\gamma ,k+q}^{\dagger }c_{\gamma ,-k}^{\dagger }+V_{k,k',q}^{2,\gamma }c_{\gamma ,k+q}^{\dagger }c_{\gamma ,k}\right)+h.c.\right]\nonumber \\
 & +\sum _{k,k',q}\left[a_{\alpha ,k'-q}^{\dagger }a_{\alpha ,k'}\right.\label{eq:hso}\\
 & \left.\times \, \left(V_{k,k',q}^{3,\gamma }c_{\gamma ,k+q}^{\dagger }c_{\gamma ,-k}^{\dagger }+V_{k,k',q}^{4,\gamma }c_{\gamma
 ,k+q}^{\dagger }c_{\gamma ,k}\right)+h.c.\right]\, \, \, ,\nonumber 
\end{alignat}
where $\gamma =x,y,z$ and the $V_{k,k',q}^{i,\gamma }$
are rather lengthy expressions given in Appendix \ref{matrixhso}.

\begin{figure}
\begin{center}\includegraphics[  width=\columnwidth,
  keepaspectratio,
  origin=c]{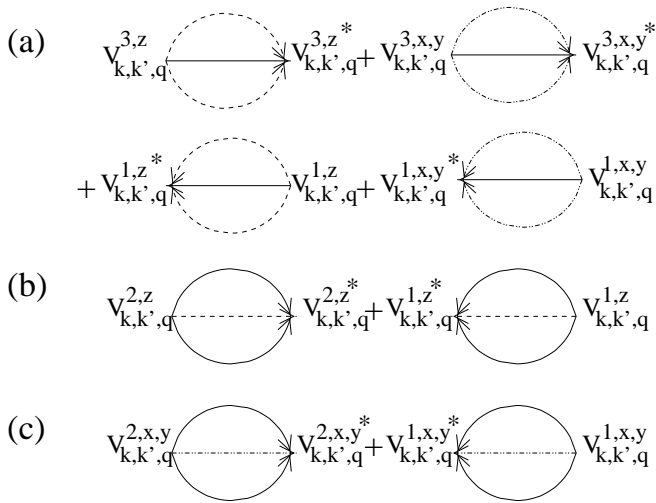}\end{center}
\caption{\label{fig2}Self energy of the spin and orbital triplet Green's
functions. (a) $\Sigma _{a_{x,y,z}}(k,\omega )$ (b) $\Sigma _{c_{z}}(k,\omega )$
(c) $\Sigma _{c_{x,y}}(k,\omega )$. Full lines denote $G_{a_{x,y,z}}^{0}(k,\omega )$,
dashed lines $G_{c_{z}}^{0}(k,\omega )$ and dotted lines
$G_{c_{x,y}}^{0}(k,\omega )$.}
\end{figure}

At the points of complete dimerization and in a pure spin system the
bond-boson description is characterized by a vanishing of the inter-dimer
coupling. This is \emph{not} the case in the spin-orbital chain,
i.e., even at the dimer-point $J_{1,2}$=3/4, the matrix elements
$V_{k,k',q}^{i,\gamma }$ do not vanish and the sum of (\ref{hnull})
and (\ref{eq:hso}) remains a non-trivial interacting problem. In
turn, spin-triplets can excite orbital-triplets by virtue of $H_{SO}$
and vice versa. For an approximate account of the resulting renormalization
of the one particle excitations, we include the effects of $V_{k,k',q}^{i,\gamma }$
perturbatively up to second order in the self energy. The one particle
excitations result form the poles of the matrix Green's function \begin{equation}
\mathbf{G}_{\nu}=\left(\begin{array}{cc}
 G_{\nu,11} & G_{\nu,12}\\
 G_{\nu,21} & G_{\nu,22}\end{array}
\right)=\mathbf{G}_{\nu}^{0}+\mathbf{G}_{\nu}^{0}\Sigma _{\nu}\mathbf{G}_{\nu}\, \, \, ,\label{eq:Dyson}\end{equation}
where the second index-pair refers to normal ($11$ and $22$) and anomalous ($12$ and $21$) propagators for particles of type $\nu$,
where $\nu = a_{x,y,z}$ or $\nu = c_{x,y,z}$ labels spin or orbital triplets.
Because $\mathbf{G}_{\nu}^{0}$ is calculated from $H_{LHP}$ it is diagonal, i.e. $G_{\nu,12}^0=G_{\nu,21}^0=0$. 
For the non-diagonal elements of $\mathbf{G}_{\nu}$ this does not apply because the matrix-self energy $\Sigma _{\nu}$ is not diagonal.
Fig. \ref{fig2} depicts all contributions to $\Sigma _{\nu}$ up to second order, where each diagram occurs with both, an incoming and an outgoing propagator on either side, i.e. such that the self energy corrections can be
written in terms of a $2\times 2$-matrix notation for each of the $3$ orbital and $3$ spin triplets. 
These diagrams show that the spin(orbital) triplet motion is renormalized
by emission of two orbital(spin) triplets. Earlier variational calculations,
performed in ref. \onlinecite{schollwoeck} have been based on a \emph{subset}
consisting of exactly these intermediate states.
\begin{figure}
\begin{center}\includegraphics[  width=0.80\columnwidth,
  origin=c]{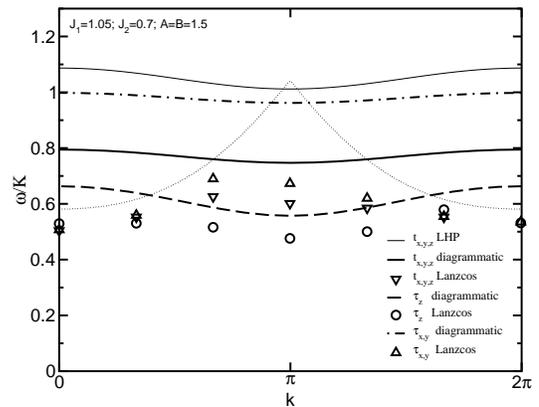}\end{center}
\vspace{-1cm}

\caption{\label{fig3}Self energy corrected triplet dispersions of the spin-orbital
chain. Full lines: spin triplets with (thick line) and without (thin
line) self-energy corrections. Dashed line: orbital $z$-triplets.
Dot-dashed line: orbital $x,y$ triplets. The dotted line show the
border of the multi-soliton continuum \cite{schollwoeck}. The symbols
denote the spin and orbital triplet energies obtained by exact diagonalization studies (Lanczos) for a system of
the length 12 (rungs) \cite{schollwoeck}. ED momenta have been back-folded into the dimerized Brillouine.}
\end{figure}
We have evaluated the self energy contributions of Fig. \ref{fig2}
numerically. The dressed Green's function spectrum displays undamped
triplet quasi-particles as well as a three-particle continuum. 
Close to the dimer point their respective spectral ranges are well
separated. In Fig. \ref{fig3} we
show the dispersion of the quasi-particles only. For comparison this
figure contains complementary data form exact diagonalization(ED)\cite{schollwoeck}.
Firstly, within a finite range of momenta around the zone center and
apart from a global relative shift of the spectra by $\sim 0.1$,
reasonable agreement can be observed for the low-energy excitations,
i.e. the $\tau _{z}$ orbital-triplets. Secondly, with respect to the
LHP, inclusion of the self-energy from Fig. \ref{fig2} improves the
agreement with ED. Thirdly, close to the zone boundary the ED spectrum refers to the lower edge of the orbital two-soliton continuum, rather than the orbital triplet, i.e. the two-soliton bound-state \cite{schollwoeck}. This two-soliton continuum is lacking in the approximate bond-operator approach.

\begin{figure}
\begin{center}\includegraphics[  width=0.85\columnwidth,
  keepaspectratio]{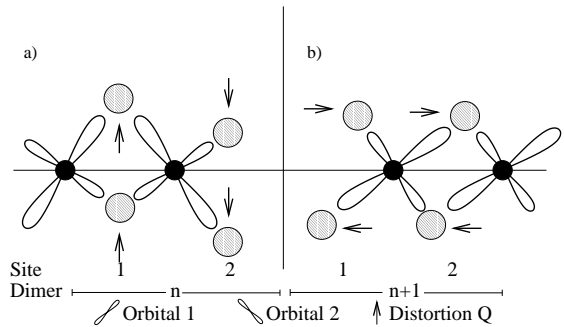}\end{center}
\caption{\label{fig4}Chosen phonon modes. a) transverse distortion leading
to antiferro-orbital ordering and b) longitudinal distortion leading
to ferro-orbital ordering. The loops labeled orbital one/two denote
the $3d_{yz}$/$3d_{zx}$ orbital respectively. The hatched circles
symbolize super-exchange orbitals of p-symmetry. Also shown is
the static distortion $Q$. The different size of the orbitals denote
the different Coulomb repulsion by electrons belonging to neighboring
atoms: the larger an orbital is drawn, the lower is its energy.}
\end{figure}

\section{Lattice dynamics}

We are now in a position to consider the phonon-renormalization due
to orbital and spin excitations. To this end two steps are required.
Firstly, the equilibrium structure of the lattice in the orbitally ordered
ground state has to be clarified. Secondly, the lattice dynamics has
to be evaluated. The first step is specific to the material. Here we focus
on a situation compatible with Na$_{2}$Ti$_{2}$Sb$_{2}$O where
the 3d$_{yz/zx}$-orbitals form the 1D chain-structures, shown in
Fig. \ref{fig4} \cite{axtell,khomskii}. Several static Jahn-Teller-type
distortions could be envisaged lifting the orbital degeneracy. In
the remainder of this paper we confine the discussion to the homogeneous
\emph{ferro}(or \emph{longitudinal})- \emph{}and \emph{antiferro}(or
\emph{transverse})-orbital static \emph{and} dynamic distortion (c.f.
Fig. \ref{fig4}a) and b)). The phonon-orbiton contribution to the
Hamiltonian for both of Fig. \ref{fig4}a) and b) reads
\begin{equation}
H_{phon}=\sum _{n}\left(Q+J_{3}\left(b_{n}^{\dagger }+b_{n}\right)\right)\left(T_{1,n}^{z}\pm T_{2,n}^{z}\right)+
\omega_{b,k}b^\dagger_k b_k,\label{hphonon}.
\end{equation} 
Both phonon modes couple to the $z$-component of the orbital pseudo-spin operator $T$, where '$-$' and '$+$' refer to the transverse, i.e. Fig. \ref{fig4}a), and longitudinal, i.e. Fig. \ref{fig4}b), phonon mode. $Q$ accounts for the static Jahn-Teller distortion, $b_{n(k)}^{\left(\dagger \right)}$
are phonon destruction(creation) operators at site(momentum) $n\left(k\right)$
and $\omega _{b,k}$ is the bare phonon dispersion.

\begin{figure}
\begin{center}\includegraphics[  width=0.80\columnwidth,
  keepaspectratio]{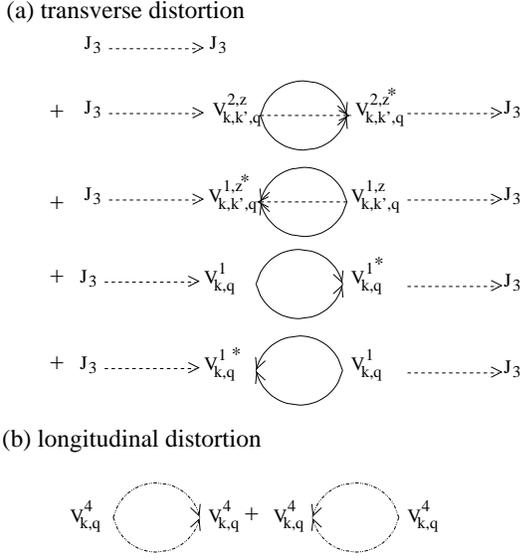}\end{center}
\caption{\label{fig5} Self energy of the phonon propagator with (a) transverse and (b) longitudinal distortion. As in Fig. \ref{fig2} full lines denote $G_{a_{x,y,z}}^0(k,\omega)$, dashed lines $G_{c_{z}}^0(k,\omega)$ and dotted lines $G_{c_{x,y}}^0(k,\omega)$. 
Note, that $G_{c_{z}}^0(k,\omega)$ is calculated with the shifted $\tilde{\tau}^{(\dagger)}_{z,k}$ (see eq. \ref{taushift}) instead of the pure $\tau^{(\dagger)}$, that was used in the calculations without any distortion.
The explicit expressions for the different $V_{k,k',q}^{i,z}$, $i=1,2$ and $V_{k,q}^{j}$, $j=1,4$ are given in appendix \ref{matrixhso}.
}
\end{figure}

\subsection{Transverse distortion}

\begin{figure}
\begin{center}\includegraphics[  width=0.75\columnwidth,
  keepaspectratio,
  angle=270,
  origin=c]{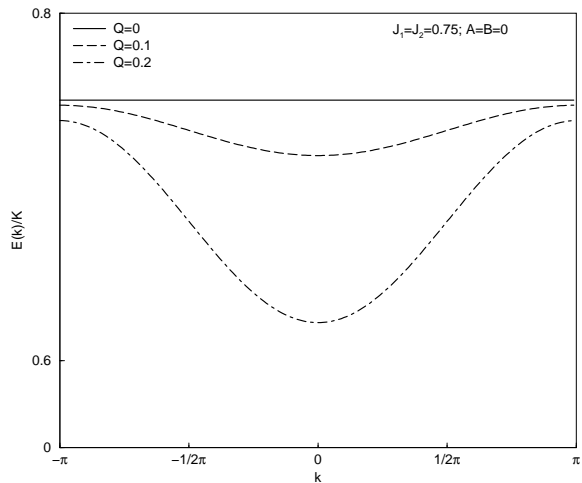}\end{center}
\vspace{-1cm}

\caption{\label{fig6}Self-energy-corrected spin triplet dispersion without
(full line) and with different (dashed and dot-dashed lines) strength of distortion $Q$
at the dimer point.}
\end{figure}

First we consider the transverse mode of Fig. \ref{fig4}a), associated
with antiferro-orbital ordering (AF-OO). In this case the electrons
on two neighboring sites occupy alternating orbitals. Expressing $T_{1}^{z}$
and $T_{2}^{z}$ in terms of bond-operators and after Fourier transformation
(\ref{hphonon}) reads: \begin{equation}
\begin{split} H_{trans}^{phon}= & \sum _{k}\underbrace{Q(\tau _{z,0}+\tau _{z,0}^{\dagger })}_{a)}\\
 + & \underbrace{J_{3}(b_{k}^{\dagger }\tau _{z,k}+b_{-k}\tau _{z,k}+h.c.)}_{b)}+\omega _{b,k}b_{k}^{\dagger }b_{k}\, \, \, ,\end{split}
\label{eq:w2}\end{equation}
 with a total Hamiltonian of $H=H_{LHP}+H_{SO}+H_{trans}^{phon}$.
The contribution linear in $\tau _{z,0}$ resulting from the static
distortion, i.e. a) in (\ref{eq:w2}), can be eliminated by realizing
that $H_{LHP}$ is quadratic in $\tau _{z,k}^{\left(\dagger \right)}$
and by invoking a constant shift of the latter operators through:
\begin{equation}
\tau _{z,k}^{\left(\dagger \right)}\rightarrow \tilde{\tau }_{z,k}^{\left(\dagger \right)}-\frac{Q\delta _{k,0}}{\left(1+J_{2}-\left(J_{2}-\frac{3}{4}+J_{2}A-\frac{3}{4}B\right)\cos k\right)}\, .\label{taushift}\end{equation}
On the level of the LHP this does not modify the $\tau _{z}$-dispersion.
However, inserting (\ref{taushift}) into $H_{SO}$ of (\ref{eq:Hso}),
new hopping terms for the spin triplets are generated modifying their
dispersion. Moreover, additional three-triplet vertices occur, because of which orbital triplets now couple to a
two-spin-triplet-continuum. These new terms read
\begin{eqnarray}
H_{SO}^{new}&=&\sum_{k,q}V_{k,q}^1 (c_{z,k}^{\dagger}a_{x,y,z,q}^{\dagger}a_{x,y,z,-k-q}^{\dagger}+\nonumber \\
&+&c_{z,-k}^{\dagger}a_{x,y,z,-q} a_{x,y,z,k+q} + h.c.)\nonumber \\
&+& \sum_{k,q}V_{k,q}^2 (c_{z,k}^{\dagger}a_{x,y,z,q}^{\dagger}a_{x,y,z,k+q} +h.c.)
\end{eqnarray}
They lead to additional self-energy contributions for the Green's function of the orbital triplets which have to be included in the phononic self-energy because of the coupling of the orbital triplets to the phonons. Explicit expressions for the
resulting spin-triplet dispersion $\tilde{\omega }_{\gamma ,k}$ on
the quadratic level, as well as the matrix elements $V_{k,q}^{i}$ with $i=1,2$ of this new
spin-orbital interaction are given in appendix \ref{transdist} and appendix \ref{matrixhso},
respectively. In Fig. \ref{fig6} we show $\tilde{\omega }_{\gamma ,k}$ for various values of $Q$ at the dimer point. As is obvious, the static Jahn-Teller distortion leads to a finite triplet dispersion.

To evaluate the phonon propagator $G_{b_k,b_k^{\dagger}}(k,\omega)$ we use the Dyson equation (\ref{eq:Dyson}) with the self energy depicted in Fig. \ref{fig5}. This includes corrections up to
second order in the phonon-orbiton-spin interactions. The phonon spectrum $S_b(k, \omega)$ is then defined as
\begin{equation}
S_b(k,\omega)= -\frac{1}{\pi} \mathrm{Im} G_{b_k,b_k^{\dagger}}(k,\omega). \label{eq:propagator}
\end{equation}
Due to the alternation of the lattice distortion in the AF-OO case, i.e. the '+'-sign in
(\ref{hphonon}), there is only a linear coupling of the phonon to
the orbital $z$-triplets in (\ref{eq:w2}). Therefore, the dominant
renormalization of the phonon-spectrum is a direct mixing between
$z$-orbitons and phonons. The remaining self-energy corrections which
are shown in Fig. \ref{fig5}(a) lead to two- and three-particle continua.

\begin{figure}
\begin{center}\includegraphics[  width=0.87\columnwidth,
  keepaspectratio,
  origin=c]{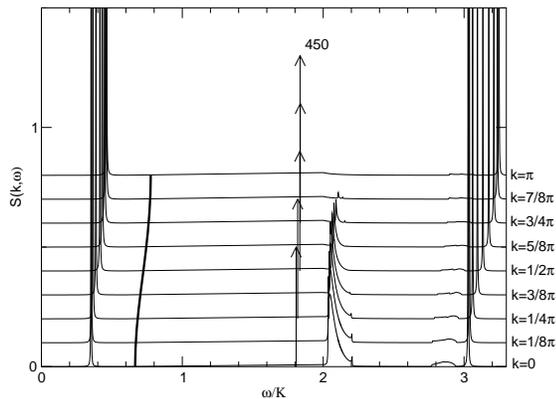}\end{center}
\vspace{-1cm}

\caption{\label{fig7}Phonon spectrum $S_b(k,\omega )$ at $J_{1}=1.2$, $J_{2}=0.8$,
$A=B=1.5$, $J_{3}=0.35$ and $\omega _{b}=1,8$ calculated with $\eta=0.001$. The left peak is
the one-orbiton excitation peak, strongly shifted to lower energies
by the distortion of the system. Note the peak position without any distortion (thick black line). 
The position of the phonon-energy
peak is shown by up-arrows. Because it is an undamped pole of the phonon Green's
function it is a $\delta $-peak, the height of which is given for
one example in the figure. Its width is set by the numerical broadening $\eta$. The two-particle continuum at $\omega \approx 2.0-2.2$
shows the two van-Hove singularities one expects for a continuum of
two non-interacting particles. At $\omega \approx 2.75-3.3$ one finds
the three-particle continuum.}
\end{figure}

Figure \ref{fig7} depicts the results of a numerical calculation
of the phonon-spectrum. 
In this figure the bare orbiton and phonon energies have been chosen to be well separated, with a phonon-energy above the orbiton
one. First, the spectrum displays two one-particle peaks which are simple poles of the phonon Green's function, i.e. the renormalized
phonon and orbiton. Their widths are set by an arbitrarily small numerical parameter $i\eta $. The bare phonon energy is momentum independent, i.e. $\omega _{b}=1.8$. This energy remains almost
unrenormalized, acquiring a very small dispersion and an upward-shift.
The energy of the orbiton-peak is clearly lowered with respect to
its bare position (thick solid line in Fig. \ref{fig7}). This shift
increases with increasing $Q$, rendering the AF-OO unstable at sufficiently
large values of the static distortion. Apart from the one-particle
peaks Fig. \ref{fig7} displays the anticipated two- and three-particle
continua. Their relative weight, as well as the momentum dependence
of their shape - which exhibits characteristic van-Hove singularities
- is determined by the matrix elements of (\ref{eq:hso}) listed in
appendix \ref{matrixhso}.

\begin{figure}
\begin{center}\includegraphics[  width=0.75\columnwidth,
  keepaspectratio,
  angle=270,
  origin=c]{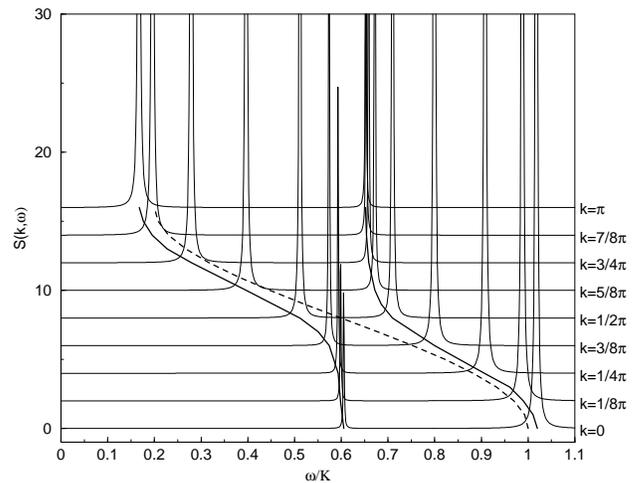}\end{center}
\vspace{-1.5cm}

\caption{\label{fig8}The phonon peak in the phonon propagator spectrum crossing
the orbiton energy at the dimer point, i.e. $J_{1}=J_{2}=0.75$,
$A=B=0$ with $\omega _{b}=(0.6+0.4\cos k)$ $Q=0.1$ and $\eta=0.001$. The dotted
line shows $\omega _{b}$ as it is set, while the thick full lines
denote the peak position of the two quasi-particles. The quasi-particle
excitations at $k\approx \pi /2$ are mixtures of orbitons and phonons
as expected for interacting particles.}
\end{figure}

For dispersive phonons, a crossing of the bare phonon and orbiton
dispersion may occur, where however phonon-orbiton coupling will lead
to level repulsion in the renormalized spectra as shown in Fig. \ref{fig8}.
In that case the quasi-particles in the vicinity of $k=\pi $/2 correspond
to an approximately equal-amplitude mixture with even and odd parity
of orbitons and phonons.

\subsection{Longitudinal distortion}

Now we consider the longitudinal mode of Fig. \ref{fig4} b), which
we associate with ferro-orbital ordering (F-OO). After a Fourier and 
Bogoliubov transformation and inserting the bond-boson expression
for $T_{1}^{z}$ and $T_{2}^{z}$ (\ref{hphonon}) reads 
\begin{eqnarray}\label{verz2}
H_{long}^{phon}&=&-2iQ\sum _{k}c_{x,k}^{\dagger }c_{y,k}-c_{y,k}^{\dagger }c_{x,k}  \\
&+&V_{k,q}^3\left(b_q^{\dagger}+b_{-q}\right)\left(c_{x,k}^{\dagger}c_{y,k+q}-c_{y,-k-q}^{\dagger}c_{x,-k}\right)
\nonumber \\
&+&V_{k,q}^4\left(b_q^{\dagger}+b_{-q}\right)\left(c_{x,k}^{\dagger}c_{y,-k-q}^{\dagger}-c_{y,k+q}c_{x,-k}\right) \nonumber.
\end{eqnarray}
Again the exact expressions for $V_{k,q}^{i}$ with $i=3,4$ are given in appendix \ref{matrixhso}. 
This leads to the diagrams shown in Fig. \ref{fig5}b) for the
longitudinal distortion. The contribution
of (\ref{verz2}) which is static with respect to the distortion lifts
the degeneracy of the orbital $x$- and $y$-triplets. This leads
to new triplet operators $c_{1\left(2\right),k}=+\left(-\right)ic_{x,k}+c_{y,k}$ with an energy-splitting of $\Delta \omega _{c}=4Q$. \textbf{\textcolor{red}{}}Yet, to simplify matters the phonon-propagator is evaluated discarding this effect, using the Bogoliubov transformed $x$- and $y$- orbital-triplets of (\ref{hquad}). 
\begin{figure}
\begin{center}\includegraphics[  width=0.70\columnwidth,
  keepaspectratio,
  angle=270,
  origin=c]{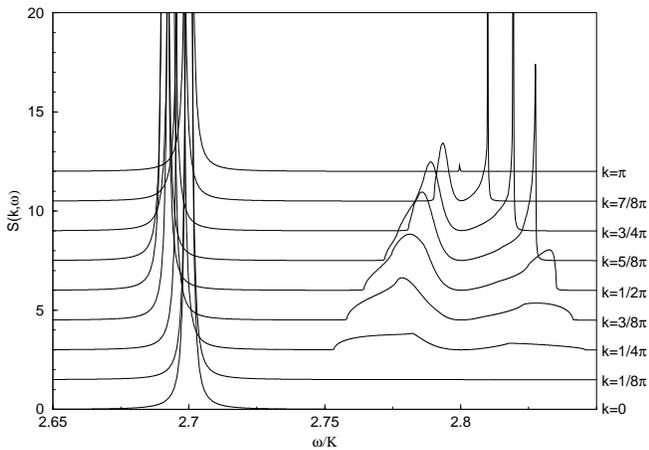}\end{center}
\vspace{-1.5cm}

\caption{\label{fig9}Spectrum of the phonon propagator in the system distorted
with the longitudinal distortion. Here the parameters are $J_{1}=1.2$,
$J_{2}=0.8$, $A=B=1.5$, $J_{3}=0.3$, $\omega _{b}=2.7$, $Q=0.1$ and $\eta=0.0001$. 
Note that there is no feature except the phonon-peak at
$k=0$. This means this kind of distortion will not be seen in spectra
of optical experiments.}
\end{figure}

While for the AF-OO a linear coupling of phonons and orbitons results
from (\ref{hphonon}), the '$-$'-sign in the latter eqn. leads to
(\ref{verz2}), i.e. the decay of phonons into orbital triplet-pairs
in the case of F-OO. In turn the primary effect on the phonon spectrum
in the F-OO is the appearance of the a two-orbiton continuum. This
can be observed in Fig. \ref{fig9} which depicts the numerical solution
of the Dyson equation of Fig. \ref{fig5} b). The spectrum shows a
renormalized simple phonon-pole the width of which is set by a numerical
broadening of $i\eta =i 10^{-4}$. The weak
phonon-dispersion observable is an interaction effects, since the
bare phonon has been chosen to be purely optical. In addition to
the renormalized phonon Fig. \ref{fig9} exhibits a two-orbiton continuum,
the spectral weight of which is more than one order of magnitude larger
than for the AF-OO case. In the case of the finite splitting $\Delta \omega_c$ included Fig. \ref{fig9} would show three superimposed continua, a case which we leave to future analysis. As for the AF-OO the continuum shows characteristic
van-Hove type singularities. Most important however, at momentum $k=0$
the spectrum resembles a pure phonon excitation only. This is caused by a vanishing of the matrix element $V_{k,q}^4$ at $q=0$ (see App. \ref{matrixhso}) and implies that this kind of distortion will not
have any effect on spectra of optical experiments e.g. Raman-scattering.

\section{conclusion}

In conclusion we have studied the interplay of spin, orbit and lattice dynamics in a model of an anisotropic spin-orbital coupled chain in the vicinity of the dimer-point. The spin-orbital sector of this model has been treated by mapping onto interacting bond-bosons with a singlet condensate and massive triplet modes. We have analyzed the lifting of the orbital degeneracy for two kinds of dynamic Jahn-Teller distortions leading either to antiferro- or ferro-orbital ordering. Studying the phonon dynamics in both cases we have found shake-up sidebands which are critically dependent on the type of distortion. In particular we have found (no) optical activity of the phonon in the transverse (longitudinal) distortion.

\noindent
{\bf Acknowledgment:} We thank C. Jurecka for stimulating discussions in the beginning of this work. This research was supported in part by the DFG, Schwerpunktprogramm 1073.
\appendix

\section{Matrix elements of $H_{0}$ \label{matrixhnull}}

For the matrix elements of the quadratic part of the Hamiltonian (Eq. \ref{hnull}) one finds $M_{n,n}^{12}=J_{1}$,
$M_{n,n}^{34}=J_{2}(1+A/2)$, $M_{n,n}^{56}=J_{2}$, $M_{n+1,n}^{12}=M_{n+1,n}^{21}=M_{n+1,n}^{11}=M_{n+1,n}^{22}=-1/4(J_{1}-3/4-1/4B)$,
$M_{n-1,n}^{34}=M_{n-1,n}^{43}=M_{n-1,n}^{33}=M_{n-1,n}^{44}=-1/4(J_{2}-3/4)$
and $M_{n-1,n}^{56}=M_{n-1,n}^{65}=M_{n-1,n}^{55}=M_{n-1,n}^{66}=-1/4(J_{2}(1+A)-3/4(1+B))$.
All the other matrix elements are zero.

\section{Spin and orbital triplet-dispersions and Bogoliubov coefficients}

\label{dispersionen} Here the spin and orbital triplet dispersions
after LHP approximation (\ref{lhpdispersion}) are given. 
\begin{eqnarray}
\omega _{a,k} & = & J_{1}\left[1-\left(1-\frac{3}{4J_{1}}(1+B)\right)\cos k\right]^{\frac{1}{2}} \nonumber\\
\omega _{c_{x,y},k} & = &
J_{2}\left[\left(1+\frac{A}{2}\right)^2-\left(1+\frac{A}{2}\right)\left(1-\frac{3}{4J_{2}}\right)\cos
k\right]^{\frac{1}{2}}\nonumber\\
\omega _{c_{z},k} & = & J_{2}\left[1-\left(1-\frac{3}{4J_{2}}(1+B)+A\right)\cos k\right]^{\frac{1}{2}}
\end{eqnarray} 
The Bogoliubov coefficients are defined as \begin{eqnarray}
u_{k}^{2} & = & \frac{1}{2}\left[1+\frac{J_{1}-\frac{1}{2}(J_{1}-\frac{3}{4}-\frac{1}{4}B)\cos k}{\omega _{\gamma ,k}}\right]\nonumber \\
g_{k}^{2} & = & \frac{1}{2}\left[1+\frac{J_{2}-\frac{1}{2}(J_{2}(1+A)-\frac{3}{4}(1+B))\cos k}{\omega _{z,k}}\right]\nonumber \\
m_{k}^{2} & = & \frac{1}{2}\left[1+\frac{J_{2}(1+\frac{A}{2})-\frac{1}{2}(J_{2}-\frac{3}{4})\cos k}{\omega _{l,k}}\right]\label{bogocoeff}
\end{eqnarray}
 and $v_{k}^{2}=u_{k}^{2}-1$, $h_{k}^{2}=g_{k}^{2}-1$, $n_{k}^{2}=m_{k}^{2}-1$.

\section{Spin-triplet dispersion for static transversal distortion}

\label{transdist} For the spin-triplet dispersion $\tilde{\omega }_{\gamma ,k}$
for the static transversal distortion one finds: \begin{equation}
\begin{split} \tilde{\omega }_{\gamma ,k}= & \left[-\frac{64Q^{2}(4(1+B)u_{k}v_{k}+(u_{k}+v_{k})^{2}\cos k)}{(7+3B-4AJ_{2})^{4}}\right.\\
 +\left(\omega _{\gamma ,k}\right.- & \left.\left.\frac{8Q^{2}(2(1+B)(u_{k}^{2}+v_{k}^{2})+(u_{k}+v_{k})^{2}\cos k)}{(7+3B-4AJ_{2})^{2}}\right)^{2}\right]^{\frac{1}{2}}.\end{split}
\end{equation}
 where $\omega _{\gamma ,k}$ is the spin triplet dispersion of the
undistorted model and $u_{k}$ and $v_{k}$ are the Bogoliubov-coefficients
(Appendix \ref{dispersionen}) of the transformations in (\ref{hquad})

\section{Matrix elements of the spin-orbital interaction $H_{so}$\label{matrixhso}}

For completeness the matrix elements of the spin orbital interaction
(\ref{eq:hso}) are listed. Here $T_{x,y}=e^{-ix}+e^{i(x-y)}$. With
$u_{k}+v_{k}=w_{k}$, $g_{k}+h_{k}=i_{k}$, $m_{k}+n_{k}=o_{k}$,
$u_{k}u_{k'}+v_{k}v_{k'}=W_{k,k'}$, $g_{k}g_{k'}+h_{k}h_{k'}=I_{k,k'}$,
$m_{k}m_{k'}+n_{k}n_{k'}=O_{k,k'}$ and $1+B=B_{1}$, $1+B/2=B_{2}$:
\begin{eqnarray}
V_{k,k',q}^{1,z} & = & -\frac{1}{8}[(2T_{k',q}+2B_{1}T_{k,-q}^{*})g_{k+q}h_{k}u_{k'-q}v_{k'}\nonumber \\
 & + & T_{k',q}(g_{k+q}h_{k}u_{k'-q}u_{k'}+h_{k+q}g_{k}v_{k'-q}v_{k'})\nonumber \\
 & + & B_{1}T_{k,-q}^{*}(g_{k+q}g_{k}u_{k'-q}v_{k'}+\nonumber \\
 & + & u_{k'}v_{k'-q}h_{k+q}h_{k})]\\
V_{k,k',q}^{1,x,y} & = & -\frac{1}{8}[(2B_{2}T_{k',q}+2T_{k,-q}^{*})m_{k+q}n_{k}u_{k'-q}v_{k'}\nonumber \\
 & + & B_{2}T_{k',q}(m_{k+q}n_{k}u_{k'-q}u_{k'}+n_{k+q}m_{k}v_{k'-q}v_{k'})\nonumber\\
 & + & T_{k,-q}^{*}(m_{k+q}m_{k}u_{k'-q}v_{k'}+\nonumber \\
 & + & u_{k'}v_{k'-q}n_{k+q}n_{k})]
\end{eqnarray}
\begin{eqnarray}
V_{k,k',q}^{2,z} & = & -\frac{1}{8}[T_{k'q}I_{k+q,k}w_{k'-q}w_{k'}\nonumber \\
 & + & 2B_{1}T_{k,-q}^{*}u_{k'-q}v_{k'}i_{k+q}i_{k}]\\
V_{k,k',q}^{2,x,y} & = & -\frac{1}{8}[B_{2}T_{k'q}O_{k+q,k}w_{k'-q}w_{k'}\nonumber \\
 & + & 2T_{k,-q}^{*}u_{k'-q}v_{k'}o_{k+q}o_{k}]\\
V_{k,k',q}^{3,z} & = & -\frac{1}{8}[2T_{k,-q}g_{k'-q}h_{k'}w_{k+q}w_{k}\nonumber\\
 & + & B_{1}T_{k',q}^{*}W_{k+q,k}i_{k'-q}i_{k'}]\\
V_{k,k',q}^{3,x,y} & = & -\frac{1}{8}[B_{2}2T_{k,-q}m_{k'-q}n_{k'}w_{k+q}w_{k}\nonumber \\
 & + & T_{k',q}^{*}W_{k+q,k}o_{k'-q}o_{k'}]\\
V_{k,k',q}^{4,z} & = & -\frac{1}{4}[(T_{k',q}+B_1T_{k,-q}^*)I_{k+q,k}W_{k',k'-q}\nonumber \\
 & + & T_{k',q}(u_{k'-q}v_{k'}+v_{k'-q}u_{k'})I_{k+q,k} \nonumber \\
 & + & B_{1}T_{k,-q}^*W_{k'-q,k'}(g_{k+q}h_k+h_{k+q}g_k)]\\
V_{k,k',q}^{4,x,y} & = & -\frac{1}{4}[(T_{k',q}B_2+T_{k,-q}^*)O_{k+q,k}W_{k',k'-q}\nonumber \\
 & + & B_2 T_{k',q}(u_{k'-q}v_{k'}+v_{k'-q}u_{k'})O_{k+q,k}\nonumber \\
 & + & T_{k,-q}^{*}W_{k'-q,k'}(m_{k+q}n_k+n_{k+q}m_k)].
\end{eqnarray}
 The matrix elements of the spin-orbital interaction $H_{so}$ arising
from the shifted orbital triplets are: \begin{eqnarray}
V_{k,q}^{1} & = & \frac{1}{4}F\sum _{k,q}\left[\cos q\cdot f_{k}\cdot (W_{k,k+q})+\right.\nonumber \\
 & + & (\cos q+\cos (k+q)+2(1+B)\cdot (1+\cos k))\times \nonumber \\
 & \times  & \left.(g_{k}u_{q}v_{k+q}+h_{k}v_{q}u_{k+q})\right]\\
V_{k,q}^{2} & = & \frac{1}{4}F\sum _{k,q}\left[2\cos q\cdot (i_{k})\cdot (u_{k}v_{k+q}+v_{q}u_{k+q})+\right.\nonumber \\
 & + & (\cos q+\cos (k+q)\cdot 2\cdot (1+B)\cdot (1+\cos k))\times \nonumber \\
 & \times  & \left.(i_{k})\cdot (W_{k,k+q})\right]\\
V_{k,q}^{3} & = & -2iO_{k,k+q} \\
V_{k,q}^{4} & = & 2i(m_k n_{k+q} - m_{k+q}n_k)
\end{eqnarray}
 where $F=Q/((J_{2}+4X)\omega _{b,0}+4J_{3}^{2})$, $X=-1/4(J_{2}-3/4+J_{2}A-3/4B)$
and $u_{k}$, $v_{k}$, $g_{k}$, $h_{k}$, $m_{k}$, $n_{k}$ are
the Bogoliubov-coefficients (Appendix \ref{dispersionen}) of the
transformations in (\ref{hquad}).

\end{document}